\documentclass[]{article}
\usepackage{lmodern}
\usepackage{amssymb,amsmath}
\usepackage{ifxetex,ifluatex}
\ifnum 0\ifxetex 1\fi\ifluatex 1\fi=0 % if pdftex
  \usepackage[T1]{fontenc}
  \usepackage[utf8]{inputenc}
  \usepackage{textcomp} % provide euro and other symbols
\else % if luatex or xetex
  \usepackage{unicode-math}
  \defaultfontfeatures{Scale=MatchLowercase}
  \defaultfontfeatures[\rmfamily]{Ligatures=TeX,Scale=1}
\fi
\IfFileExists{upquote.sty}{\usepackage{upquote}}{}
\IfFileExists{microtype.sty}{% use microtype if available
  \usepackage[]{microtype}
  \UseMicrotypeSet[protrusion]{basicmath} % disable protrusion for tt fonts
}{}
\makeatletter
\@ifundefined{KOMAClassName}{% if non-KOMA class
  \IfFileExists{parskip.sty}{%
    \usepackage{parskip}
  }{% else
    \setlength{\parindent}{0pt}
    \setlength{\parskip}{6pt plus 2pt minus 1pt}}
}{% if KOMA class
  \KOMAoptions{parskip=half}}
\makeatother
\usepackage{xcolor}
\IfFileExists{xurl.sty}{\usepackage{xurl}}{} % add URL line breaks if available
\IfFileExists{bookmark.sty}{\usepackage{bookmark}}{\usepackage{hyperref}}
\hypersetup{
  hidelinks,
  pdfcreator={LaTeX via pandoc}}
\usepackage{longtable,booktabs}
\usepackage{etoolbox}
\makeatletter
\patchcmd\longtable{\par}{\if@noskipsec\mbox{}\fi\par}{}{}
\makeatother
\IfFileExists{footnotehyper.sty}{\usepackage{footnotehyper}}{\usepackage{footnote}}
\makesavenoteenv{longtable}
\providecommand{\tightlist}{%
  \setlength{\itemsep}{0pt}\setlength{\parskip}{0pt}}
\usepackage{graphicx}
\usepackage{hyperref}
\usepackage{pdflscape} % for 'landscape' environment
\usepackage{multirow}
\usepackage{setspace}
\usepackage{lineno}
\usepackage{authblk}
\usepackage{soul}
\usepackage{caption}

\usepackage{natbib}
\usepackage{pdfpages}
\usepackage{fancyhdr}

\title{Integration of biophysical connectivity in the spatial optimization of coastal ecosystem services\footnote{© 2020. This manuscript version is made available under the CC-BY-NC-ND 4.0 license \href{http://creativecommons.org/licenses/by-nc-nd/4.0/}{http://creativecommons.org/licenses/by-nc-nd/4.0/}}}

\author[1]{Andres Ospina-Alvarez\footnote{Corresponding author: Andrés Ospina-Alvarez, email: aospina.co@me.com; address: Spanish Scientific Research Council, Mediterranean Institute for Advanced Studies (IMEDEA-CSIC/UIB), C/ Miquel Marques 21, CP 07190 Esporles, Balearic Islands, Spain.}}

\author[2]{Silvia de Juan Mohan}

\author[3, 4]{Katrina J. Davis}

\author[5]{Catherine González}

\author[6]{Miriam Fernández}

\author[6]{Sergio A. Navarrete}

\affil[1]{Mediterranean Institute for Advanced Studies IMEDEA (UIB-CSIC), C/ Miquel Marques 21, CP 07190 Esporles, Balearic Islands, Spain.}
\affil[2]{Institute of Marine Sciences ICM (CSIC), Passeig Marítim de la Barceloneta, 37-49, 08003, Barcelona, Spain.}
\affil[3]{Department of Zoology, University of Oxford, Zoology Research and Administration Building, 11a Mansfield Road, Oxford, OX1 3SZ, United Kingdom}
\affil[4]{Australian Research Council Centre for Excellence for Environmental Decisions, The University of Queensland, St Lucia, Queensland, 4072, Australia.}
\affil[5]{Instituto de Fomento Pesquero (IFOP), Almte. M. Blanco Encalada 839, Casilla 8-V, Valparaiso, Chile.}
\affil[6]{Núcleo Milenio – Centro de Conservación Marina; Estación Costera de Investigaciones Marinas, Pontificia Universidad Católica de Chile, Alameda 340, C.P. 6513677, Casilla 193, Correo 22, Santiago, Chile.}

\begin{document}
% \linenumbers	
	\maketitle
	
	\textbf{Running page head:} Spatial optimization of coastal eco-services

% \begin{doublespacing}
	
\begin{abstract}
		\item Ecological connectivity in coastal oceanic waters is mediated by dispersion of the early life stages of marine organisms and conditions the structure of biological communities and the provision of ecosystem services. Integrated management strategies aimed at ensuring long-term service provision to society do not currently consider the importance of dispersal and larval connectivity. A spatial optimization model is introduced to maximise the potential provision of ecosystem services in coastal areas by accounting for the role of dispersal and larval connectivity. The approach combines a validated coastal circulation model that reproduces realistic patterns of larval transport along the coast, which ultimately conditions the biological connectivity and productivity of an area, with additional spatial layers describing potential ecosystem services. The spatial optimization exercise was tested along the coast of Central Chile, a highly productive area dominated by the Humboldt Current. Results show it is unnecessary to relocate existing management areas, as increasing no-take areas by 10\% could maximise ecosystem service provision, while improving the spatial representativeness of protected areas and minimizing social conflicts. The location of protected areas was underrepresented in some sections of the study domain, principally due to the restriction of the model to rocky subtidal habitats. Future model developments should encompass the diversity of coastal ecosystems and human activities to inform integrative spatial management. Nevertheless, the spatial optimization model is innovative not only for its integrated ecosystem perspective, but also because it demonstrates that it is possible to incorporate time-varying biophysical connectivity within the optimization problem, thereby linking the dynamics of exploited populations produced by the spatial management regime.
\end{abstract}

\begin{figure*}
	\centering
	\includegraphics[width=1\linewidth]{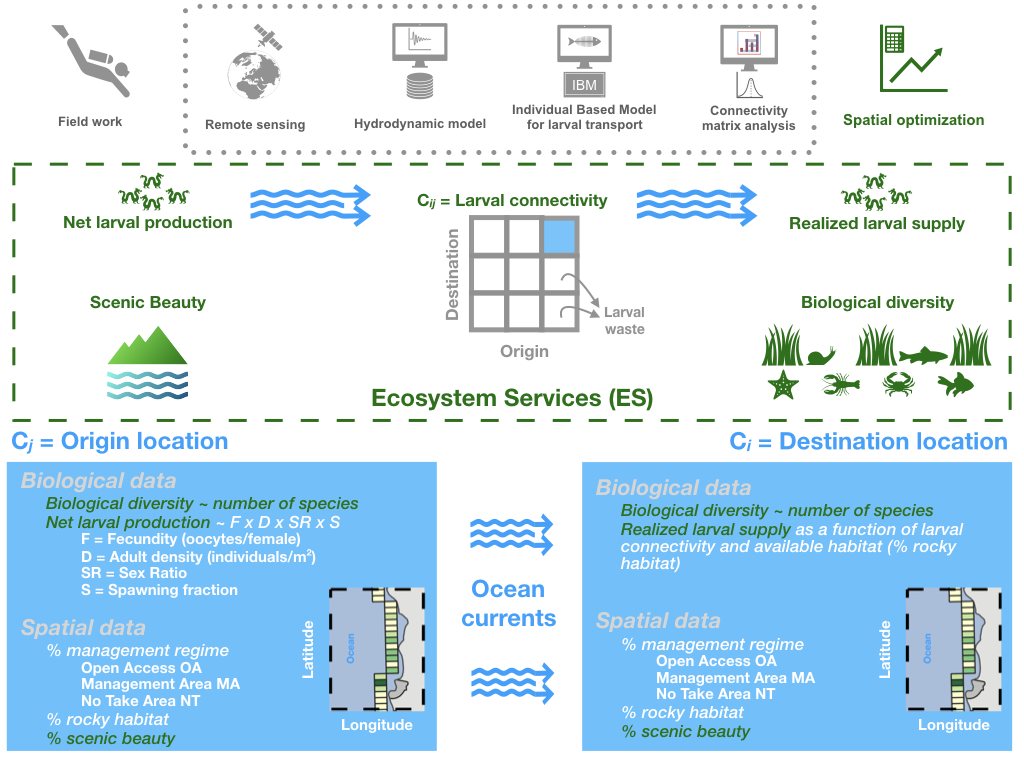}
	{\caption*{Graphical abstract.}}
	\label{fig:graph_abstract}
\end{figure*}


\section{Introduction}
Coastal marine ecosystems are being perturbed, fragmented, polluted, and subjected to wholesale biodiversity loss due to a multiplicity of human drivers, including the over-exploitation of many species of commercial interest \citep{Worm:2006ir, Crain:2009il}. Addressing this global phenomenon requires effective implementation of sustainable management practices that consider not only fished species, but also the multiple ecosystem services provided by marine and coastal areas \citep{Hutton:2003ef,Adams:2004km}. The debate over optimal strategies to manage and preserve marine ecosystems remains unresolved \citep{Hilborn:2017du, Roberts:2017ir, Worm:2006ir}. Central to the debate is the use of marine protected areas (MPA) as critical instruments for ecosystem conservation and spatial management. In fact, the increased implementation of MPAs in the last decade \citep{Jantke:2018bt} has been driven by international agreements to protect 10\% of Economic Exclusive Zones by 2020 \citep{UNEP:2010uj}. However, there are increasing concerns that the excessive focus on the areal component of the 10\% protection target (CBD Aichi target 11; \url{http://www.cbd.int/sp/ targets/}) is detrimental to an effective management \citep{Gill:2017hc} of representative and well-connected networks of MPAs \citep{Visconti:2019ci}. Spatial planning must explicitly consider inherent dependencies between marine patches under different conservation and exploitation regimes to ensure population, genetic, community and/or ecosystem flows or connectivity \citep{Smith:2018ee}.

Spatial connectivity is key for the structure of marine ecosystems, as it underlies biological as well as biophysical processes that determine ecosystem productivity, dynamics and resilience \citep{Gaines:2010dw,Watson:2012eh,Hidalgo:2017gl}. There are different types and scales of ecological spatial connectivity \citep[e.g., population connectivity, genetic connectivity, ecosystem connectivity; see][]{Brown:2016hp}, but, in general terms, ecological spatial connectivity refers to biological and biophysical processes that connect areas over comparatively large spatial scales, i.e., those that surpass the scale of spatial management instruments and which can be crucial for the persistence of marine populations and communities \citep{Carr:2017hz}. For demersal and benthic marine organisms, whose adults are predominantly sedentary, large-scale connectivity is mediated through the dispersal of early-life stages (i.e., eggs and larvae). Egg and larval connectivity patterns are driven by biophysical processes characterized by strong temporal and spatial variability \citep{OspinaAlvarez:2018hn}. Such variability modulates the structure and function of benthic ecosystems, and it is therefore an essential input for the configuration of MPA networks \citep{Fovargue:2017jg}. However, spatially and temporally dynamic connectivity among marine populations is generally ignored by spatial planning schemes, including MPA design criteria (\citealt{Leslie:2005de,Balbar:2019by,Ramesh:2019hj,Hidalgo:2019hk}; but see, \citealt{Krueck:2017dm}). Most marine conservation planning approaches focus on "potential connectivity", statically represented through the size and spatial placement of protected area units \citep[e.g.,][]{Smith:2018ee}. These approaches assume that spatial units are self-sustained in the long-term \citep{Pressey:2007dm,Magris:2015io}. Recent models \citep[e.g.,][]{White:2014crb} incorporate larval connectivity parameters, such as self-recruitment and network centrality, into spatial optimization approaches \citep[e.g., MARXAN and MARXAN-Connect;][]{Rossi:2014cj, Dubois:2016fj,Daigle:2020gc} to identify sites most relevant for meta-population persistence \citep{Magris:2015io,Krueck:2017dm}. 

An additional challenge in the science of MPA network design \citep[\textit{sensu}][]{Gaines:2010dw} is to determine how different configurations of protected areas guarantee not only biological population persistence, but also maximise the long-term provision of ecosystem services to ensure society well-being \citep{Curtin:2010km,Bennett:2015de}. "Uses" and "activities" are familiar concepts for marine resource managers, but, beyond direct economic returns, the non-material benefits associated with a given use are commonly overlooked when making decisions about protected areas \citep{Cornu:2014hj}. The incorporation of societies' values into conservation proposals \citep{MartinLopez:2012ko,deJuan:2017hc,Ban:2019eu} has been slow in terrestrial and marine habitats worldwide \citep{Daily:2009je,Bennett:2015de}. The implementation of scientifically-based decision making that considers feedback between ecosystems and social well-being remains a major challenge \citep{Congreve:2019gf, Norstrom:2020gj}. An additional challenge is posed by the generally small spatial scales of management schemes, given that ecosystem services are not always demanded in the same geographic location as the ecological processes that support them \citep{vanJaarsveld:2005fa,Rodriguez:2006fa}. The biophysical connectivity between areas is key to link ecosystems and society at regional scales \citep{Popova:2019fs,Potts:2014bf}. For example, biological communities produce larvae that can migrate long distances to recruit and replenish fished populations \citep{OspinaAlvarez:2015ia}, providing resources to local fishers. Also, carbon storage by seagrasses benefits society at a global scale, and thus, the beneficiaries might be very distant from the ecosystem service providers \citep{Congreve:2019gf}. These processes are not embedded in current spatial planning practices that often overlook the open nature of marine ecosystems \citep{Maxwell:2020is}, e.g., by establishing artificial frontiers in the coastal habitat continuum. Accounting for ecological connectivity across management units at large spatial scales is therefore crucial to ensure the provision of multiple ecosystem services in the long term \citep{Potts:2014bf}.

To address these multi-disciplinary challenges, we developed a spatial optimization model that explicitly incorporates the inherent dynamism of coastal biophysical systems \citep{OspinaAlvarez:2018hn}. Ecological connectivity is estimated through coupled biophysical models of egg and larval dispersal along a coastal system. This approach differs from most decision-making tools by embedding larval connectivity matrices in the optimization algorithm, rather than incorporating connectivity as a stationary property of coastal metapopulations  \citep[e.g., MARXAN;][]{Chollett:2017dp}. The optimization model was built based on a case study in the central coast of Chile, a highly productive region within the Humboldt Upwelling Ecosystem, where wind-driven coastal upwelling supports productive industrial and small-scale fisheries \citep{Thiel:2007tt,Chavez:2009dw}. A circulation model is used to capture realistic dynamics of the coastal ocean in this area, this model has previously been applied to examine dispersal and connectivity of coastal invertebrates \citep{OspinaAlvarez:2018hn,Blanco:2019fe}. Biological data from the coastal ecosystem dominated by intertidal and subtidal rocky habitats, the habitat of the principal commercial species for small-scale fisheries in the region, is also available \citep{deJuan:2015db}. The management regime for the small-scale fisheries follows a territorial user rights for fisheries (TURF) spatial scheme, locally known as Areas de Manejo y Explotación de Recursos Bentónicos \citep[AMERBs;][]{Fernandez:2005gk,Gelcich:2012dm}. The available data set also includes the demand for ecosystem services by users of the coast in this area \citep{deJuan:2017hc}. 

The model prioritises marine zoning to achieve ecosystem conservation and social benefits in areas with productive coastal ecosystems that supply a set of services demanded by coastal users \citep{deJuan:2015db}. Locally, the present work is timely because the well-established TURF system has been the subject of heated debate in the Chilean Congress following approval of The National Biodiversity Strategy 2017 – 2030 in 2018 (Decree 14, Ministry of Environment, Chile). The strategy mandates the use of coastal planning approaches that incorporate CBD recommendations for the implementation of protected areas  \citep{MinisteriodelMedioAmbienteChile:2018wt}. Globally, the proposed model offers a promising practical tool to design and implement large-scale integrated management of connected coastal social-ecological systems.

\section{Methods}
\subsection{Model Domain}
\begin{figure}
	\centering
	\includegraphics[width=1\linewidth]{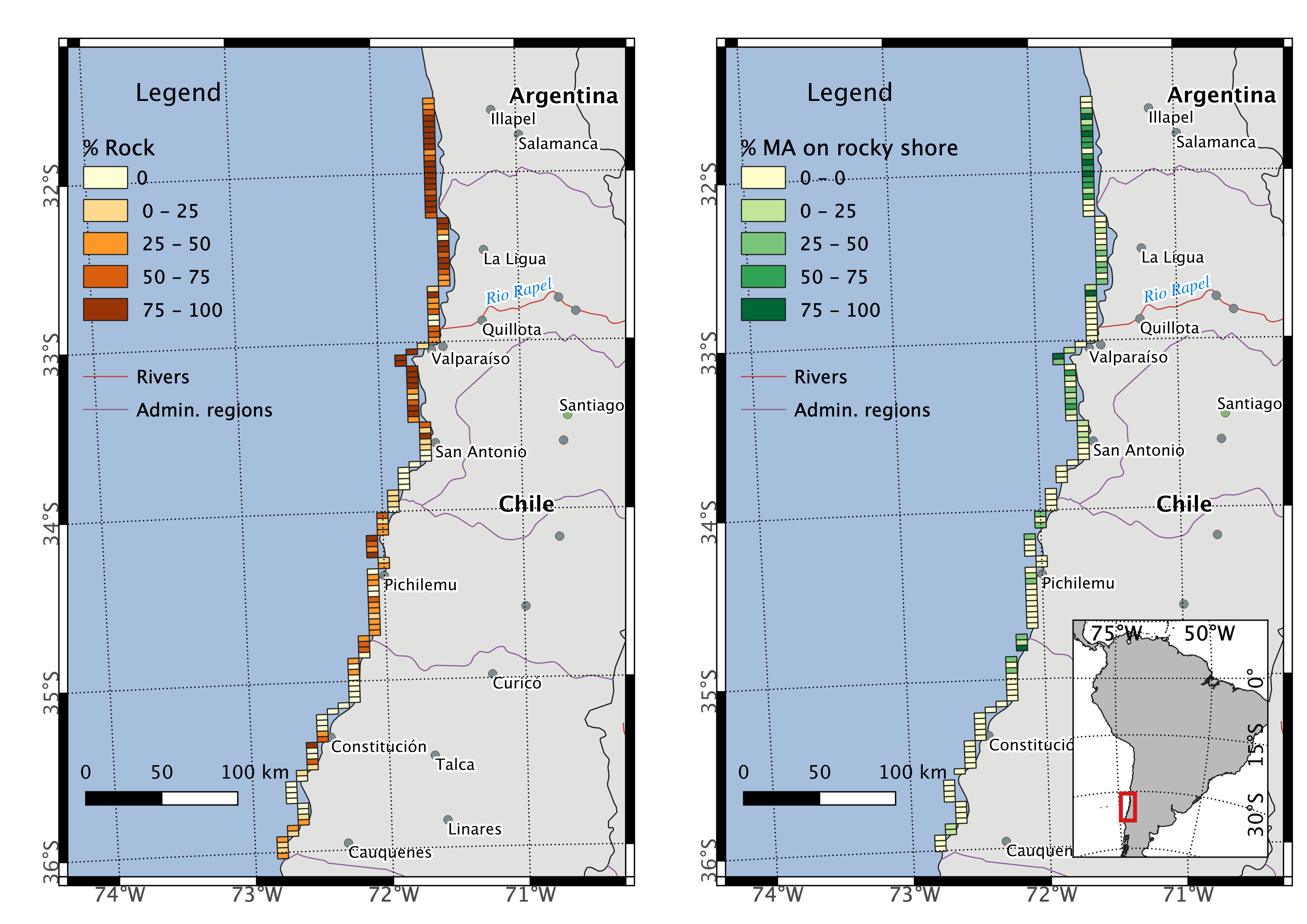}
	\caption[Study area]{Study area maps showing (left) the percentage of rocky shore substrate per unit area and (right) the percentage of managed areas (MA on rocky shore substrate), including TURFs, Natural Monuments and Natural Sanctuary areas.}
	\label{fig:study_area}
\end{figure}

The model domain extends along the central coast of Chile (31.57ºS to 36.00ºS), with a spatial resolution of 2 latitudinal km (Fig.\ref{fig:study_area}). It corresponds to the Eastern Boundary Upwelling Ecosystem of the south-east Pacific \citep{Strub:1998ut,Thiel:2007tt}. This ecosystem is dominated by the south-east Pacific atmospheric anticyclone, which provides increasingly steady upwelling-favourable winds to the north, and strong but seasonally variable winds to the south. The main surface circulation feature is the Humboldt Current, which flows northwards, approximately 200 km off the coast. A more coastal surface current (Chilean coastal current) also flows predominantly to the north, and responds to upwelling-favourable wind forcing \citep{Aiken:2008ih}. This coastal surface current is essential for the transport of coastal-released gametes \citep{OspinaAlvarez:2018hn}.

Within the model domain there are 75 operative TURFs; 1 coastal Natural Monument and 4 coastal Natural Sanctuary areas, covering \emph{ca.} 31\% of the shoreline. Data describing the spatial distribution of these protected areas (hereafter MA, e.g., Managed Areas) was obtained from the Undersecretary of Fisheries (SUBPESCA) and the Ministry of the Environment (Ministerio del Medio Ambiente in Spanish; Fig.\ref{fig:study_area}, right panel). Digitalization of the spatial data was conducted with the software QGis v3.8. Rocky shore areas (e.g., areas dominated by rocky vs. sandy substrate) were the target habitat in this study as it sustains the main coastal fisheries in the region. Habitat distribution was obtained by characterizing the habitat type (rock vs. sand) in each of the latitudinal cells using high resolution satellite photography (Fig.\ref{fig:study_area}, left panel) for the intertidal zone and projecting this onto a uniform shallow subtidal band down to a depth of 200 m. Combined, these two data sources describe the proportion of rocky coast covered by MA in each cell. The remaining fraction of rocky coast in each cell was assumed to be open access \citep[hereafter OA; see][for details]{Blanco:2017je}.

\subsection{Biological data}
Shallow subtidal rocky habitats along the central Chilean coast are characterized by kelp forests (\emph{Lessonia trabeculata}) and a diverse assemblage of mobile macro-invertebrates, rock fishes, birds and marine mammals that predominantly predate or graze benthic species \citep{PerezMatus:2017eh}. Several of these rock fish and invertebrate species are subjected to intense exploitation, as divers or coastal gatherers mostly target the muricid gastropod, ‘loco’, exported as the Chilean abalone (p1, \emph{Concholepas concholepas}), key-hole limpets (p2, \emph{Fissurella spp.}) and red sea urchin (p3, \emph{Loxechinus albus}).

Biological surveys to estimate the net larval production (\emph{L}) of the three most important commercial species, loco, key-hole limpet and red sea urchin, were conducted in the subtidal zone in 2012-2014. Four sites were selected within the model domain to collect samples and assess the adult size and gonad investment of key-hole limpets and red sea urchins \citep[see][for details]{Blanco:2017je}. Data collected by the research group were used to estimate adult size and gonad investment of Chilean abalone (n = 2,900). An additional set of surveys obtained in seven sites within this domain, including MA, OA and no-take areas (hereafter NT), were used to assess adult density \citep[D; see][for details]{deJuan:2015db}. These variables (per capita gonad production and adult density) allowed us to estimate potential fecundity (F) and, from this, net larval production (L), which was estimated from net egg production assuming no mortality during egg development \citep[see eqs. 1-5 in][]{OspinaAlvarez:2015ia}. Based on these studies \citep[see][for details]{Blanco:2017je,Blanco:2019fe}, we estimated that net larval production in MAs was 17\% less than in NT areas, which was the baseline (Table \ref{table:gametes}). Scale factors to estimate L in open access areas were calculated for each species from field work and laboratory experiments \citep[see][]{Blanco:2017je}. Afterwards, L per cell was estimated based on potential fecundity per unit of rocky area for the different fishing regimes (see \citealt{Blanco:2017je} and eqs. from 2 to 8 in \citealt{Blanco:2019fe}). Additionally, average benthic species richness ($ V_z $) observed in the rocky subtidal habitat was collected by SCUBA dive surveys of benthic communities, i.e. sub-canopy sessile benthos and mobile invertebrates, and reef fish in the seven sampling sites, with paired MA and OA. Records of average number of species under each fishing regimen provided estimates of an average of 10.2 species per m\textsuperscript{2} in OA, 11.7 in MA, and 12.1 in NT (see details on the methodology in \citealt{deJuan:2015db}).

\begin{longtable}[]{@{}llll@{}}
\caption{Scale factor for larval production (L) per species for each management regime. No-take management regime areas were considered as the baseline.} \label{table:gametes} \\
\toprule
\begin{minipage}[t]{0.22\columnwidth}\raggedright
\strut
\end{minipage} & \begin{minipage}[t]{0.22\columnwidth}\raggedright
Open Access Areas (OA)\strut
\end{minipage} & \begin{minipage}[t]{0.22\columnwidth}\raggedright
Managed Areas (MA)\strut
\end{minipage} & \begin{minipage}[t]{0.22\columnwidth}\raggedright
No-Take Areas (NT)\strut
\end{minipage}\tabularnewline
\begin{minipage}[t]{0.22\columnwidth}\raggedright
Loco (p1)\strut
\end{minipage} & \begin{minipage}[t]{0.22\columnwidth}\raggedright
0.230\strut
\end{minipage} & \begin{minipage}[t]{0.22\columnwidth}\raggedright
0.830\strut
\end{minipage} & \begin{minipage}[t]{0.22\columnwidth}\raggedright
1\strut
\end{minipage}\tabularnewline
\begin{minipage}[t]{0.22\columnwidth}\raggedright
Limpet (p2)\strut
\end{minipage} & \begin{minipage}[t]{0.22\columnwidth}\raggedright
0.126\strut
\end{minipage} & \begin{minipage}[t]{0.22\columnwidth}\raggedright
0.830\strut
\end{minipage} & \begin{minipage}[t]{0.22\columnwidth}\raggedright
1\strut
\end{minipage}\tabularnewline
\begin{minipage}[t]{0.22\columnwidth}\raggedright
Sea urchin (p3)\strut
\end{minipage} & \begin{minipage}[t]{0.22\columnwidth}\raggedright
0.350\strut
\end{minipage} & \begin{minipage}[t]{0.22\columnwidth}\raggedright
0.830\strut
\end{minipage} & \begin{minipage}[t]{0.22\columnwidth}\raggedright
1\strut
\end{minipage}\tabularnewline
\bottomrule
\end{longtable}

\subsection{Biophysical connectivity}
The ocean velocity fields were obtained from the HYbrid Coordinate Ocean Model (HYCOM) analysis \citep{Chassignet:2007jh}, a data assimilating forecast of the global ocean circulation with sufficiently high resolution to accurately reproduce mesoscale processes that dominate ocean variability. Then, a Spatially Explicit Individual Based Model (SEIBM) was used to simulate the dispersal of early life stages of the three commercial species. The SEIBM was coupled with the 3D hydrodynamic model using a customized version of the open source modelling tool ICHTHYOP \citep{Lett:2008fo}. The code was used to simulate trajectories of early life stages from velocity fields in a 3D hydrodynamic model, using Pelagic Larval Durations (PLD) of the three model species: 90 days for the Chilean abalone, 20 days for the red-sea urchin and 10 days for key-hole limpets. During this development time, larvae are subjected to advective and diffusive processes that condition their alongshore, cross-shore and vertical movement in the water column and, consequently, the total "larval waste" and successful onshore recruitment at the end of the development period. In consequence, larvae leaving the studied domain through the northern, southern and western open boundaries and larvae advected into oceanic waters are considered to be lost for recruitment. We considered that larvae display Diel Vertical Migration (DVM) behaviour, previously shown to effect recruitment rates onshore but which does not significantly alter spatial or temporal patterns of connectivity \citep{OspinaAlvarez:2018hn}. A connectivity score that incorporates seasonality was used to capture observed reproductive peaks of the model species: summer and winter for \emph{Concholepas}, and spring and summer for key-hole limpets and the red sea urchin. The connectivity score represents the standardized recruitment number of particles transported from spawning to recruitment locations weighted by demographic and habitat data. The spawning locations were assumed to be dependent only on available habitat and were therefore identical for all species. The optimal number of particles to be released in order to reach stable patterns was established at 20,000 per day, every five days (i.e., 120,000 particles per month). For a detailed description of the methodology see \cite{OspinaAlvarez:2018hn}.

A "connectivity matrix" (C; as defined by \citealt{Cowen:2009fm}), per dispersal event was generated. Potential connectivity was defined as the probability of larval transport from a spawning site \emph{j} (columns in C) to a destination location \emph{i} (rows in C), after completing development. Thus, the diagonal of C is the probability of local larval retention, and the sub-diagonals indicate the probability of dispersal between sites. The C matrix contains all the necessary information for metapopulation dispersal dynamics and time-varying processes \citep{Aiken:2014cx}: the character of the ocean dictates that it is typically time-varying and asymmetrical. Realized larval connectivity corresponds to larval supply and it can be defined as the effective number of larvae that travel from \emph{j} to \emph{i} and the corresponding matrix ($ H_{jip} $) can be estimated for each species (\emph{p}) using potential connectivity matrices weighted by relevant biological and environmental information \citep{Watson:2010fa}. Specifically, habitat availability (percentage of rocky and sandy coastline), management regime (percentage of MA, NT and OA) and adult abundance and size (as a factor determining egg and larval production) per cell were used to weight the potential matrix. All these factors were considered invariant over the timescales of dispersal.

\begin{figure}
	\centering
	\includegraphics[width=1\linewidth]{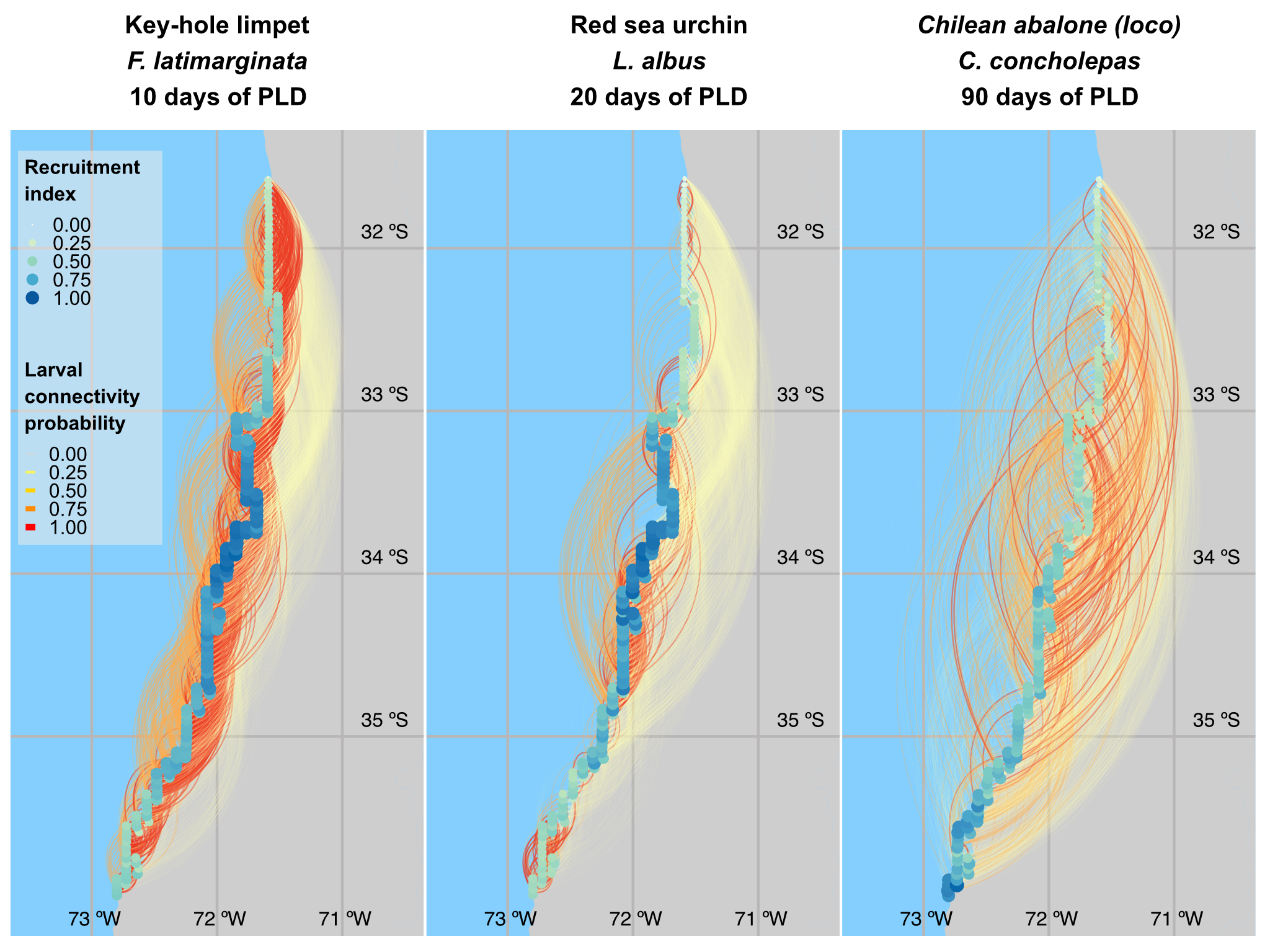}
	\caption{Larval connectivity networks for key-hole limpet (p2), red sea urchin (p3) and \emph{C. concholepas} (p1). The connection between two locations is represented as curved lines (arcs) in a yellow-red colour scale that represents larval connectivity probability. Recruitment intensity is represented by dots (green-blue colour scale). The larval connectivity probability is extracted from C matrices and the recruitment index represents the standardized number of recruited larvae at each location. The figure is based on patterns and results described by \cite{OspinaAlvarez:2018hn}.}
	\label{fig:connect}
\end{figure}

In consequence, the model was structured into six main components:

\begin{itemize}
\tightlist
\item
  environmental envelope: proportion of rocky coast and prevailing
  management regime $ z\; \in \; \left( OA,\; MA,\; NT \right)\ $.
\item
  source sites (i.e., \emph{j}, spawning sites).
\item
  destination sites (\emph{i}, settlement sites).
\item
  net larval production (L) as a function of density, size and fecundity
  of adult individuals.
\item
  potential larval connectivity, C: probability that larvae spawned at
  site, \emph{j}, will end up at site \emph{i}.
\item
  realized larval supply (H): C matrix multiplied by the number of
  larvae released in cell \emph{j} which is a function of the larval per
  capita production, the environmental envelope of the source cell
  (binary) and the abundance of adults in that cell. The management
  regimen does not play a role as microscopic larvae are transported throughout
  the study site (Fig.\ref{fig:connect}).
\end{itemize}

\subsection{Quantification of ecosystem services}
The prioritisation of ecosystem attributes in the study area was assessed through face-to-face interviews with the principal end-users \citep{deJuan:2017hc}. Over 900 questionnaires were conducted in six coastal locations within the model domain, where small-scale fisheries and national tourism are the principal activities; therefore, these questionnaires targeted local fishers, residents and tourists. The survey was designed to elicit user groups’ prioritisation of ecosystem attributes. Despite socioeconomic variability among the study locations, ecosystem valuation by users, and among users’ groups, was highly homogeneous. Several attributes were relevant to all users, like scenic beauty and clean waters and beaches; however, fishers also prioritised fishing resources and biodiversity, while tourists and residents prioritised intangible values including the peace and relaxing atmosphere of the coast \citep[see more details on the approach in][]{deJuan:2017hc}. 

The potential of coastal areas to provide these services was estimated from bio-physical data in the area. The highly valued "scenic beauty" of the coastline was identified by users as the absence of urban development on rocky coasts \citep{deJuan:2017hc}. This attribute can also be indirectly related with clean water, and peace and relaxation, the other intangible services valued by users. The proportion of the coast within the model domain that has not been modified by urban development was identified through Google Earth images. This information was overlapped with the characterization of the coast as rocky substrate and provided the proportion of each cell characterized by the attribute scenic beauty. The biological diversity and fishing resources, mainly prioritised by fishers, are also are included in our model.

The connectivity matrices C, described in the previous section, define which areas make disproportionately large contributions to realized larval supply to other areas along the coast. Sites that excel for their net larval production ultimately supply adult populations in the region and are key to maintaining populations of commercial species in the area, as long as they themselves receive enough individuals to maintain the local population. Therefore, biological productivity in the region is assumed to be well-correlated with net larval production at a given site, since this variable integrates adult abundance and size. Based on its link with biological productivity, net larval production was used as a proxy for the potential of a site to secure  ''food production'' (i.e., cath of fishing resources) as an ecosystem service.

The average number of benthic species ($ V_z $) estimated for the coastal areas under the different protection regimens was used as an additional input variable in the optimization model, when the target was to maximise the biological diversity of a site (see next section for more details) as the baseline for the provision of numerous ecosystem services, including recreational value \citep[as identified in][]{deJuan:2017hc}.

\subsection{Spatial optimization model}
We developed a spatial optimization model to identify the optimal configuration of marine zoning (e.g. locations of MA, NT and OA in the study area) that maximises the potential of coastal areas to meet integrated ecosystem management targets. Our modelling framework builds on previous approaches \citep[e.g.,][]{Davis:2015gv, Watts:2009wi}, but uses positive decision variables with linear programming versus binary decision variables (using, e.g. mixed integer programming). All models are solved in GAMS (General Algebraic Modelling System, v.23.5.2), available at \url{https://www.gams.com/}, using the CPLEX solver \citep{CPLEX:2009vw}. We considered three targets, which aim to maximise the potential provision of different sets of ecosystem services. In target 1 (T1), we maximise net larval production (as a proxy for food production). In target 2 (T2), we maximise net larval production and biodiversity. In target 3 (T3), we maximise net larval production, biodiversity, and scenic beauty. Spatial priorities consistent with target 1 require identifying the best spawning \textit{j} sites in terms of net larval production, weighted by the proportion of larvae successfully arriving to any site \emph{i}, for each commercial species. Biodiversity (T2 and T3) is represented through benthic species richness. To increase biodiversity in the study region, spatial zoning should prioritise zones that support higher levels of benthic species richness, e.g., NT. Finally, scenic beauty (T3) values are increased by locating NT areas in parts of the coast with high scenic beauty values, e.g., cells where > 80 \% of total area is rocky habitat and there is  no urban development.

For each target, we assessed a range of scenarios based on existing or potential management regimes in central Chile. Scenario 1 (S1) is our baseline scenario, and describes the existing management regime in central Chile. This scenario provides a baseline against which we explored the potential for alternative management scenarios to increase (or decrease) the potential provision of ecosystem services to society. Scenarios 2 and 5 describe pessimistic and optimistic (S2 and S5 respectively) outcomes for potential ecosystem service provision in the study region. Scenario 3 is based on realistic projections of management changes that could take place in the region. Finally, scenario 4 incorporates the objectives of the CBD targets by considering the assignation of 10\% of the coast to no-take areas based on a spatial configuration that aims to maximise biological connectivity constrained by holding an integral approach that explicitly considers multiple ecosystem attributes. 

\begin{itemize}
\tightlist
\item	
Scenario 1 (S1): Maintains existing management conditions.
\item
Scenario 2 (S2): Final allocation is constrained to 100 \% OA areas. This is a
worst-case scenario from the point of view of conservation and
sustainability.
\item
Scenario 3 (S3): Model is constrained by existing ratio of MA and OA areas
(30:70) but is free to allocate these areas in space.
\item
Scenario 4 (S4): Model solution is constrained to include existing MA, but freely allocates 10\% of the rocky study area to NT (from existing OA areas).
\item
Scenario 5 (S5): Final allocation is constrained to 100 \% NT areas. i.e. there
is loss of an ecosystem service (fishery), but it achieves the maximum
possible conservation.
\end{itemize}

The characteristics of each scenario constrain the integrated ecosystem management target that can be achieved. Net larval production can be maximised (or assessed) across all five scenarios, therefore we assess the optimal configuration of the study region to meet target 1 for each scenario. However, for targets 2 and 3 we consider only a subset of scenarios. To achieve target 2, e.g., maximise net larval production and biodiversity, we assess scenarios 3-5. Scenarios 1-2 are not assessed for this target because they contain no NT zones, therefore outcomes would repeat the outcomes of target 1. Target 3 maximises net larval production, biodiversity and scenic beauty. As scenic beauty is maximised by allocating a cell with rocky habitat, and >80 \% of its area free from urban development, to NT, we limit analysis for this target to scenario 4. 

To identify the optimal spatial configuration of the study area that meets each integrated ecosystem management target, we define three objective functions. The constraints applied to each function are specific to the five scenarios. We use linear programming with positive decision variables ($ X_{iz} $) to identify the optimal allocation of each cell \emph{i} to each management zone \emph{z} (MA, NT, OA). Decision variable $ X_{iz} $ will take a positive value if cell \emph{i} is allocated to management regime \emph{z}, and 0 otherwise. The following constraint (eq. 1) ensures that each cell is fully allocated across the three zones, but not over-allocated (e.g. > 1).
\begin{equation}\label{eq.1}
\sum_{iz}^{}{X_{iz}\; =\; 1}
\end{equation}
The objective function for target 1 (e.g. maximise net larval production) incorporates realized larval supply and the probability that a larvae will settle in a given cell. As defined previously, realized larval supply ($ H_{jip} $), describes potential larval connectivity, the number of larvae released from a destination cell (\emph{j}), and other qualities of the destination cell, including the environmental envelope and abundance of adults of a given species (\emph{p}). The probability that a larvae will settle in a given cell is determined by settlement probability $ \lambda_{ipz} $, which is specific for each species \emph{p} and each management regime \emph{z}, and the environmental envelope of the destination site, i.e., the percentage of available rocky habitat in each destination cell \emph{i}. We maximise the following objective function (Eq. 2) to achieve target 1:
\begin{equation}\label{eq.2}
\max \sum_{ipz}^{}{X_{iz}\; H_{jip}\; \lambda _{ipz}}
\end{equation}
In scenarios 3 and 4, this objective function is subject to an additional constraint, which ensures that 30 \% of the rocky area ($R_i$) in the study region is allocated to the MA zone (S3), and 10 \% of rocky area is allocated to the NT zone (S4):
\begin{equation}\label{eq.3}
\sum_{iz}^{}{R_{i}}\;\times \; M_{z}\; =\; \sum_{iz}^{}{X_{iz}}R_{i}
\end{equation}
Specifically, this constraint (Eq. 3) specifies that the rocky area in each cell allocated to a specific management regime, must be equal to $ M_z $, where $ M_{MA} $ = 0.3 for scenario 3, and $ M_{NT} $ = 0.1 for scenario 4.

The objective function for target 2 builds on the objective function of target 1, including all general (Eq. 1), and scenario-specific constraints (Eq. 3), but includes a term to describe the diversity value $ V_z $ of each management regime \emph{z} (eq. 4). 
\begin{equation}\label{eq.4}
\max \sum_{ipz}^{}{X_{iz}\; H_{jip}\; \lambda _{ipz}\; V_{z}}
\end{equation}
The objective function for target 3 is the same as target 2 but subject to an additional constraint: that cells allocated to the NT zone must have at least 80\% of their rocky area free from human construction.

We normalised the following $ x\; \in \; \left( H_{jip},\; V_{z} \right) $ ecosystem service values between 0 and 1 following the equations below (eqs. 5-7), so that realized larval production and diversity values were comparable and contributed equally to the optimization value.

\begin{equation}\label{eq.5}
a \; =\; min(x)
\end{equation}

\begin{equation}\label{eq.6}
b \; = \; max(x)
\end{equation}

\begin{equation}\label{eq.7}
x_{n}\; =\; \frac{\left( x-a \right)}{\left( b-a \right)}
\end{equation}

\section{Results}

\begin{figure}
	\centering
	\includegraphics[width=1\linewidth]{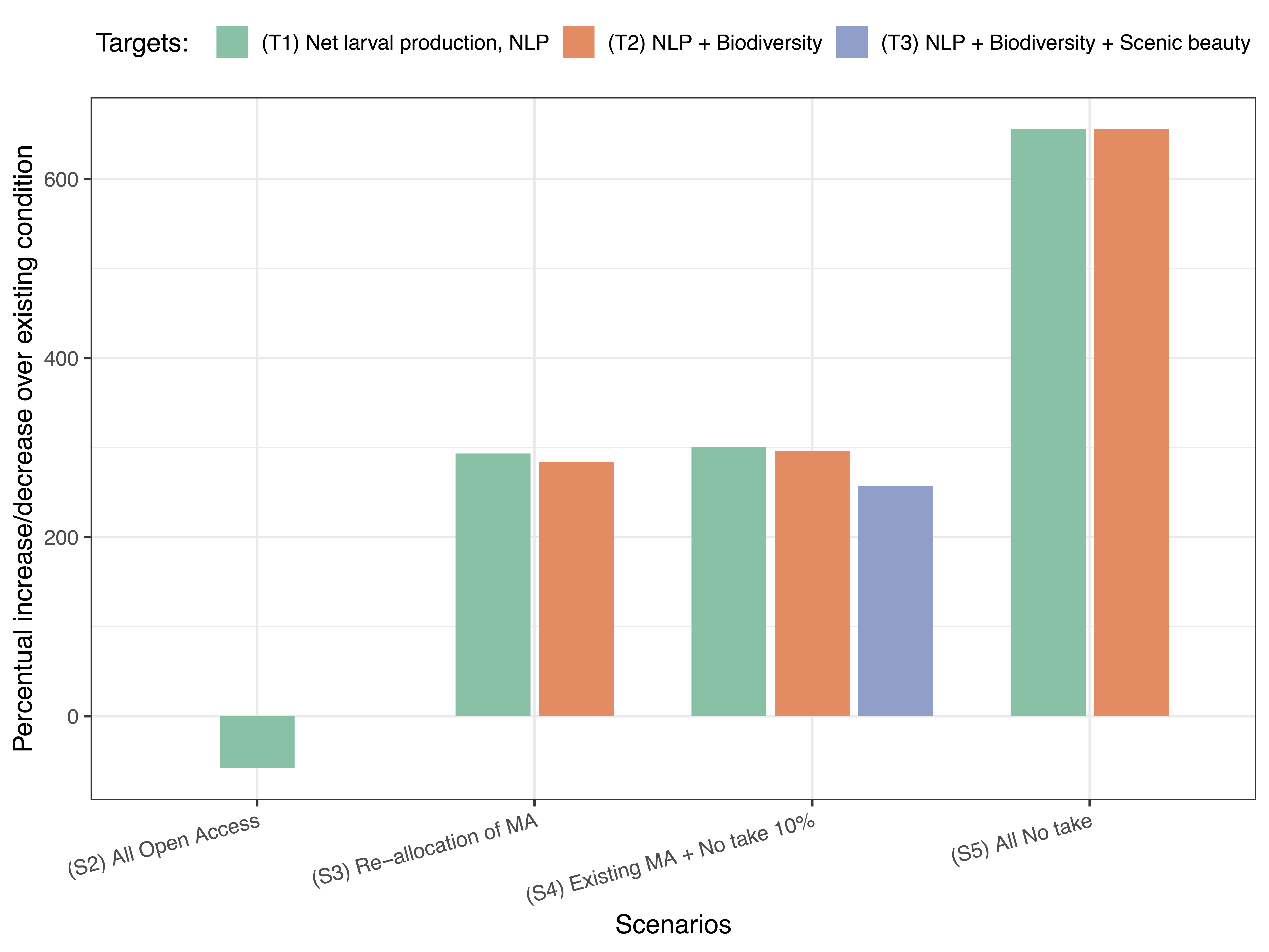}
	\caption{Optimization results for all possible combinations of targets and scenarios. Notice that some target-scenario combinations are not possible or have the same value as scenarios evaluated under Net larval production (NLP, T1) because for NLP + Biodiversity (T2) and NLP + Biodiversity + Scenic beauty (T3) species diversity ($ V_z $) is an average value for each type of management area \emph{z} (MA, NT, OA) and scenic beauty is a constraint only affecting the allocation in NT for T3.}
	\label{fig:optim_results_barplot}
\end{figure}

\begin{figure}
	\centering
	\includegraphics[width=1\linewidth]{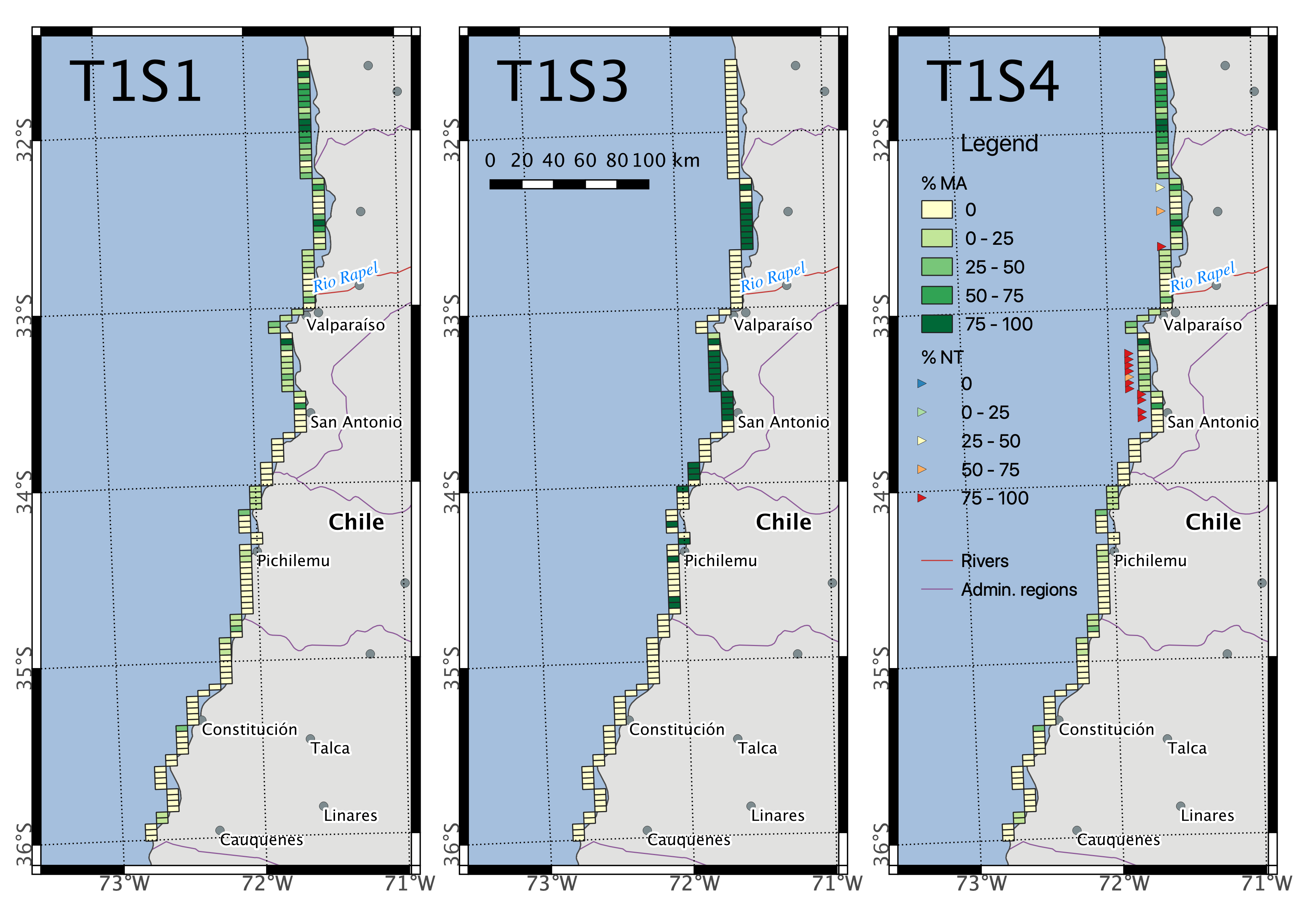}
	\caption{Scenarios to allocate management or/and no take areas from the optimization model. Net larval production -- existing MA (T1S1); Net larval production --  Existing MA proportion but optimal allocation (T1S3); and  Net larval production -- Existing MA allocation, but additional 10\% of study area changed from OA to NT zones (T1S4). The legends for \% MA and geographic features and the scale are shared by all panels.}
	\label{fig:optim_results_1}
\end{figure}

Net larval production slightly increased under the existing management scenario in central Chile (S1), compared to the worst-case scenario with 100\% of the shore allocated to OA areas (S2; Table \ref{table:results}). However, by allowing the optimization model to freely allocate the spatial location of the existing MA and OA areas (S3), i.e. without altering total area (and therefore stock) proportions, the objective function (17.49) increased relative to S1 and S2 (Table \ref{table:results} or Fig.\ref{fig:optim_results_barplot}). A similar objective function value (17.94) was obtained when an additional 10\% of open access area was re-allocated to no-take area (S4), without altering the existing geographic distribution of MA. These scenarios (S3 and S4) represent a 3- and 5-fold improvement over the current and worst-case scenarios, respectively. However, under these conditions, the management areas (MA) and the no-take areas (NT) are highly overlapped and concentrated in the central and northern sections of the model domain, which may be undesirable (Fig.\ref{fig:optim_results_1}, panel T1S3 and T1S4). As expected, allocating the entire shore to no-take areas (S5) produced the highest net larval production (objective function: 39.08). It is important to emphasize that net larval production is being optimized by embedding the connectivity matrices in the optimization algorithm. Therefore, a site with low connectivity or with low self-recruitment will not get a high priority in the optimization exercise, even if that site had a high net larval production. 

\begin{table*}
\caption{Targets, scenarios and optimization values. Net larval production values are described in Eqs. \ref{eq.2} and \ref{eq.4}), species diversity optimization value corresponds to $ V_z $ in (Eq. \ref{eq.4}). The objective function is the resulting output from the optimization equation and constraints applied to each target-scenario combination. Consider that diversity ($ V_z $) for T2 and T3 is an average value for each type of management area \emph{z} (MA, NT, OA) and that scenic beauty is a constraint only affecting the allocation in NT for T3. In consequence, some target-scenario combinations are not possible or have the same value as scenarios evaluated under T1.}
\label{table:results}
\resizebox{\textwidth}{!}{%
\linebreak
\begin{tabular}{@{}lllrrr@{}}
	&  &  & & & \\
\cline{1-2} \cline{3-6}
\textbf{Target number} & \multicolumn{1}{c}{\textbf{}} & \multicolumn{1}{c}{\textbf{}} & \multicolumn{3}{c}{\textbf{Optimization value}} \\ \cline{4-6} 
\textbf{and description} & \multicolumn{1}{c}{\textbf{Scenario number}} & \multicolumn{1}{c}{\textbf{Scenario description}} & \multicolumn{1}{c}{\textbf{\begin{tabular}[c]{@{}c@{}}Larval\\ connectivity\end{tabular}}} & \multicolumn{1}{c}{\textbf{\begin{tabular}[c]{@{}c@{}}Species\\ diversity\end{tabular}}} & \multicolumn{1}{c}{\textbf{\begin{tabular}[c]{@{}c@{}}Objective\\ function\end{tabular}}} \\ \hline

\multirow{5}{*}{\begin{tabular}[c]{@{}c@{}}1\\ Net larval production (NLP)\end{tabular}} & 1 & Existing & 5.96 & - & 5.96 \\
&  &  & \multicolumn{1}{l}{} & \multicolumn{1}{l}{} & \multicolumn{1}{l}{} \\
& 2 & 100\% allocated to OA & 3.45 & - &  3.45\\
&  &  & & & \\
& 3 & \multicolumn{1}{l}{\begin{tabular}[c]{@{}l@{}}Existing MA proportion but optimal\\ allocation [MA: 0.3, OA: 0.7]\end{tabular}} & 17.49 & - & 17.49 \\
&  &  & & & \\
& 4 & \begin{tabular}[c]{@{}l@{}}Existing MA allocation, but additional\\ 10\% of study area changed from\\ OA to NT zones [MA: 0.3, OA: 0.6, NT: 0.1]\end{tabular} & 17.94 & - & 17.94 \\
&  &  & & & \\
& 5 & 100\% allocated to NT & 39.08 & - & 39.08\\
&  &  & & & \\
\hline

\multirow{3}{*}{\begin{tabular}[c]{@{}c@{}}2\\ Net larval production + Biodiversity\\ (NLP + Biodiversity) \end{tabular}} & 3 & \multicolumn{1}{l}{\begin{tabular}[c]{@{}l@{}}Existing MA proportion but optimal\\ allocation [MA: 0.3, OA: 0.7]\end{tabular}} & 17.49 & 51.21  & 16.95 \\
&  &  & & & \\
& 4 & \begin{tabular}[c]{@{}l@{}}Existing MA allocation, but additional\\ 10\% of study area changed from\\ OA to NT zones [MA: 0.3, OA: 0.6, NT: 0.1]\end{tabular} & 17.74 & 51.63 & 17.65 \\
&  &  & & & \\& 5 & 100\% allocated to NT & 39.08 & 55.99 & 39.08 \\
&  &  & & & \\
\hline

\begin{tabular}[c]{@{}c@{}}3\\ Net larval production +\\ Biodiversity + Scenic beauty\\ (NLP + Biodiversity + Scenic beauty)\end{tabular} & 4 & \begin{tabular}[c]{@{}l@{}}Existing MA allocation, but additional\\ 10\% of study area into\\ NT (area taken from OA) [MA: 0.3, OA: 0.6, NT: 0.1]\end{tabular}  & 15.43 & 51.63 & 15.33 \\
\hline
\end{tabular}
}
\end{table*}

\begin{figure}
	\centering
	\includegraphics[width=1\linewidth]{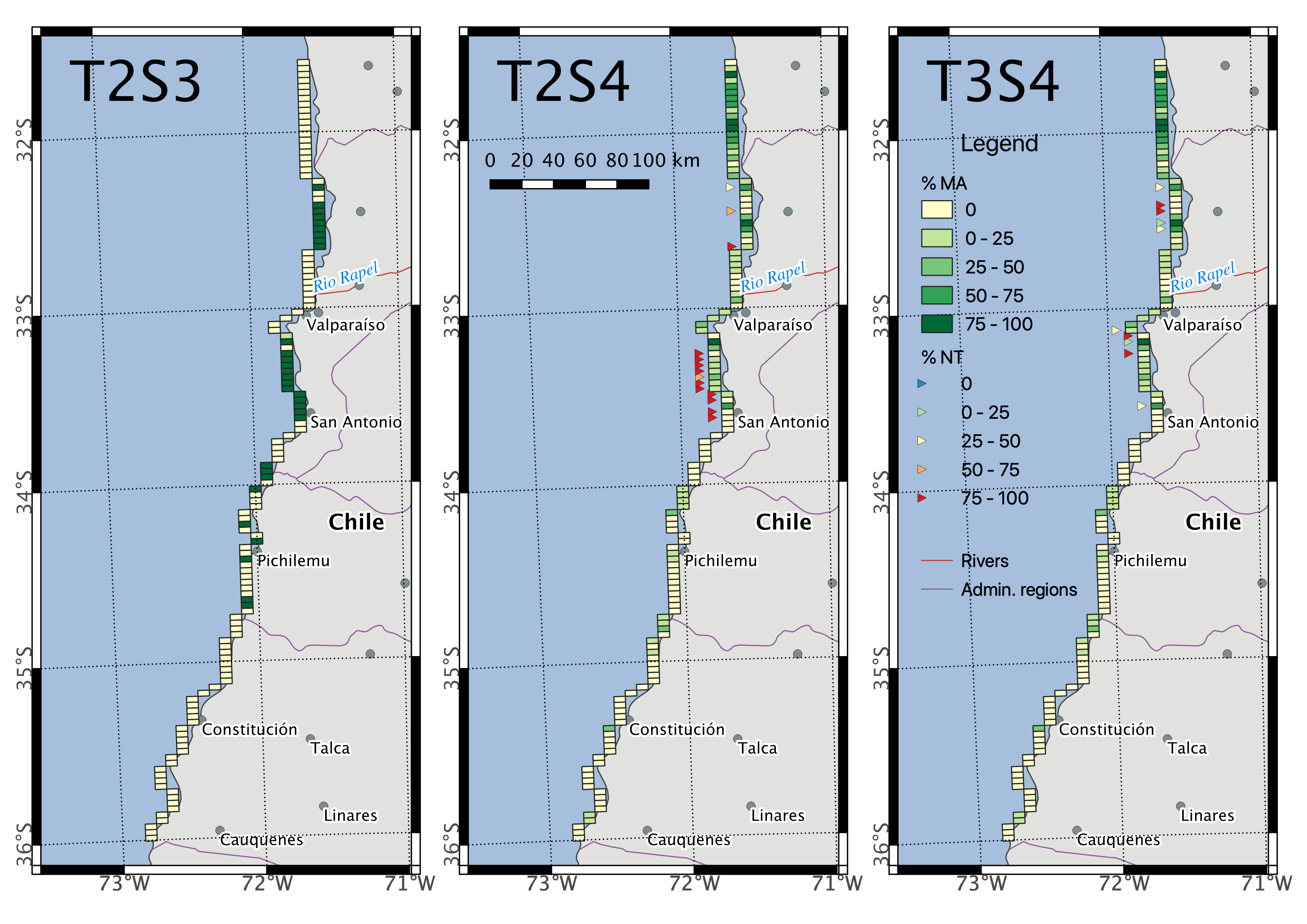}
	\caption{Scenarios to allocate management or/and no take areas from the optimization model. Net larval production + Biodiversity -- Existing MA proportion but optimal allocation (T2S3); Net larval production + Biodiversity -- Existing MA allocation, but additional 10\% of study area changed from OA to NT zones (T2S4); and Net larval production + Biodiversity + Scenic beauty -- Existing MA allocation, but additional 10\% of study area changed from OA to NT zones (T3S4). The legends for \% MA and geographic features and the scale are shared by all panels; the legend for \% NT is shared by T2S4 and T3S4 panels.}
	\label{fig:optim_results_2}
\end{figure}

In scenarios 3 and 4, the addition of biological diversity to net larval production (T2) obtained a similar objective function value as target 1 (i.e., only considering net larval production) (Table \ref{table:results}). This result suggests that the simultaneous maximisation of diversity and net larval production was not detrimental to net larval production in this ecosystem (Fig.\ref{fig:optim_results_barplot}), probably because MA are located in productive areas with high biological diversity. In consequence, scenarios 3 and 4 for target 1 and target 2 are very similar and most NT are allocated to cells with a proportion of MA, resulting in numerous small NT (Fig.\ref{fig:optim_results_2}, panel T2S4).

When we incorporate all the variables prioritised by end-users, by the inclusion of scenic beauty into conservation targets (target 3), and we increase 10\% of no-take area, there was a slight decrease in optimization value, with respect to target 2 (T2S4), from 17.65 to 15.33 (Table\ref{table:results}). This is a consequence of constraining the best areas for biodiversity and larval connectivity to the presence of "scenic beauty". In this scenario there was an improved spatial representativity of NT, with less overlap of NT and MA as observed in T2S4, as cells with 100\% of OA are allocated to NT (Fig.\ref{fig:optim_results_2}, panel T3S4). Note that all these scenarios provide larger optimization functions than the existing scenario.

\section{Discussion}
The spatial optimization model introduced in this work addresses the challenge of achieving a diversity of management objectives, ranging from conservation of coastal biodiversity, to the maximisation of net larval production and larval connectivity of key commercial species for local fisheries, and to societal preferences for the scenic beauty of the coast. The application of this novel integrated approach to the central coast of Chile is promising, and highlights that an additional 10\% protection of the coast could increase 5-fold the potential for ecosystem services' provision. The spatial optimization targets were founded on the basis that the fisheries management regimen (MA, NT, OA) will change the population density and size of commercial species, and the biological diversity. These changes affect the realized larval supply (proxy of biological productivity) along the coast mediated by biophysical connectivity. The different targets provided similar outputs, likely driven by correlation between variables, e.g., biodiversity and larval connectivity or biodiversity and scenic beauty provided by non-urbanized areas, which is a consequence of exploring the value of coastal areas already shaped by human intervention. Despite limited variability in the overall output (i.e., the objective function), the simultaneous achievement of multiple conservation objectives (target 3) implied a different geographical distribution of protected areas compared to the current scenario or to target 1 (when a single objective was considered), with large sections of OA areas assigned to NT, resulting in lower number of NT. This solution might be more feasible than small NT distributed over larger number of cells, which might imply high costs in surveillance. Regardless of the target, an optimal scenario implied maintaining the current location of MA but the addition of 10\% of the coast to NT. Thus, along our modelled coastline, it would be possible to achieve integrated conservation objectives and reduce the perceived negative effects of MPA implementation \citep{Ban:2019eu} at no additional cost for fishers in terms of the removal of users' rights in existing management areas. Still, fishers opposition might be detected for the closing of large OA areas (as in target 3), whereas smaller NT adjacent to MA (as in targets 1 and 2) could be supported by local fishers as part of their fisheries management plan. It is important to note that a case-by-case approach must be followed to explore alternatives of no-take area positioning that minimizes other perceived negative impacts on end-users \citep{Ban:2019eu}. 

The optimization exercise constrained by a target that incorporated scenic beauty along the coast (i.e., non-urbanized areas as a proxy for scenery, clean beaches and peaceful environments, as prioritised by end-users in the area, \citep{deJuan:2017hc}) illustrates the multi-dimensional nature of coastal ecosystem services and the feasibility of simultaneously including a set of very different ecosystem services in management decisions \citep[e.g., provision, regulating and cultural services,][]{MEA:2005wx}. Despite the simultaneous inclusion of three conservation targets causing a decrease in the objective value, it further improved the spatial representation (spacing) of protected areas (NT and MA) along the coast. It must be noted that the spatially uneven distribution of restricted areas across scenarios is a limitation of the model, as it prioritises regions where the rocky habitat for the commercial species is dominant. Previous studies have criticized the lack of analysis assessing the extent to which the representativity of many sensitive habitats has been achieved within global MPAs \citep{Fischer:2019jx,Jantke:2018bt}. Our current approach aims to improve habitat connectivity and representativity in a network of fishery restricted areas that are connected through the strong hydrodynamism of the Humboldt current \citep{OspinaAlvarez:2018hn}. However, the model was designed for the rocky coast, meaning that the central region was consistently prioritised as a conservation area, as the southern section of the model domain is characterized by greater proportions of sandy coast. This result emphasizes that habitat representativeness is an important criterion for reserve network design. Further model development should incorporate key species and diversity values from the sandy subtidal areas in the region, so habitat diversity is considered. This model extension would also allow incorporation of additional societal preferences in the central coast of Chile, linked with the diversity of coastal habitats, e.g., clam fisheries, beach tourism, surf, resulting in a truly integrative spatial optimization approach.

The positioning of management areas was highly consistent across scenarios and there is probably an historical explanation for this, as fishers tend to select the most productive sites for the allocation of management areas. Moreover, these productive sites, with strong larval connectivity, are probably also high diversity sites as other species would also match the connectivity patterns of the key species. Therefore, some management areas currently located in the best places to provide ecosystem services could be important ancillary conservation instruments if they were well enforced. The level of enforcement is key, as previous studies in central Chile observed that densities of economically important macro-invertebrates and reef fish are not significantly different between highly enforced management areas and no-take MPAs \citep{Gelcich:2012dm}. These observations are critical as from the twenty official MPAs in Chile, only five are effectively managed \citep{Mora:2006de,Petit:2018cu}. During the last decade, Chile has made some advances in the establishment of protected areas. However, a pending task is to consolidate an MPA network in the coastal area, particularly in the central region where human-uses are concentrated and no-take areas would probably face fishers opposition \citep{Suman:1999ct,Gelcich:2009gy}. In this context, the current exercise aimed to develop a tool that could assist prioritisation of management efforts for the location of protected areas and the identification of management areas that could incorporate a no-take section and/or an area to be managed under special regimes to increase enforcement level. Priority sites for conservation frequently overlapped with highly populated areas, which include large commercial harbours and polluted areas associated with hydrothermal plants (e.g., central prioritised cells in the model overlap with the urban area of Valparaiso). This overlap implies an unfeasible establishment of protected areas and highlights the need to consider all critical components of coastal ecosystems in the spatial optimization and increase spatial resolution of the information and models. With the incorporation of scenic beauty in the optimization exercise, we largely overcome this issue as, by prioritising sites that hold higher scenic beauty, it indirectly selects less urbanized sites. These local issues, that complicate spatial planning approaches, are common to many countries that lack a legal framework for the spatial management of the coastal and marine continuum. In these scenarios, the identification of priority sites relying on multiple and interacting data sources is essential for decision making.

By incorporating ecological connectivity in the spatial prioritisation of marine resources, we show that the effects of management efforts can propagate beyond the region through biophysical processes that support the provision of ecosystem services by coastal areas. Continuous improvement of physical and biological information is desirable and necessary before any geographically-specific implementation of model results. For instance, higher-resolution information on biological community structure, population size distribution and diversity, including habitat diversity (e.g., subtidal sands or kelp forests), is necessary to capture small-scale biological spatial variability. Currently, the model considers average biological metrics on "rocky shores" per management regime, shaped by the dynamism of a mid-resolution biophysical model that creates patterns of connectivity. After the connectivity modelling results, sensitivity analyses are also necessary to evaluate the different causes of larval losses and compare them with the dynamics of larval recruitment under natural conditions. In this work, we found that losses of larvae to the northern and southern boundaries and offshore western were moderate (57.6\% SD ± 21.3). However, most of them were related to oceanic advection, while losses to the North were low and occurred mainly in the austral summer and especially for long PLDs. On the other hand, losses due to beaching were moderate (39.3\% SD ± 27.3). Total mortality, the sum of all sources of losses, was 81.7\% SD ± 19.2. The low percentages of losses towards the northern and southern boundaries suggest that the high density of recruitment in the central zone of the study area is not an artefact of the larval transport model (i.e., SEIBM) but it is related to the representativeness of the rocky habitat. The central zone of the study area is the zone with the highest percentage of rocky habitat per cell. In our model, a condition for larval recruitment is the existence of rocky habitat, simulating the selection of settlement habitat by organisms living in natural conditions. An additional issue that needs to be resolved is the incorporation of temporal variability in the optimization exercise. The operational model on which implementation is based should deal with temporal variability in the hydrodynamic connectivity at scales relevant for management. 

Our model was restricted to the rocky subtidal and the values linked to this coast, which can exclude social-economic important activities such as beach tourism or clam fisheries. Future model developments should encompass the diversity of coastal ecosystems and human activities to inform a truly integrative spatial management. To achieve this, research efforts in the area should focus on acquiring biophysical data at continuous spatial scales as, for example, another model limitation was biodiversity values included as static variables linked to the presence or absence of fisheries restriction. We urge the compilation of spatially-explicit databases to obtain more precise estimates of the critical processes that sustain social-ecological systems in coastal areas. Through sensitivity analysis, the framework proposed here can be used to identify the type of information and the geographic location where new data would most contribute to improve model results. The broad adoption of ecosystem service science in coastal management is limited by the paucity of data \citep{Carcamo:2013fz,Saunders:2016bl}, therefore, models such as ours are designed to motivate the acquisition of critical information. These models should work in an adaptive way by gradually improving with the incorporation of new knowledge. The simultaneous achievement of multi-dimensional objectives is a global and critical issue for the spatial planning of human activities on the coasts, and our modelling approach offers a path in this direction. A wider application of the model includes the creation of short-, medium- and long-term projections for policy makers to promote the sustainable use of coastal ecosystems more effectively at a regional scale. The model could easily accommodate IPCC projections of sea-surface temperature and ocean currents to forecast larval growth and dispersal pathways coupled with shared social-economic scenarios to analyse feedback between climate change and social well-being \citep{ONeill:2014ep}. Moreover, the visualisation of optimal configurations of management units in maps could also assist an effective communication with relevant stakeholders. 

\section*{Author Contributions}
AO, SdJ and KD conceived the ideas and designed methodology; AO, SdJ, MF and CG contributed to data; AO, KD and SN performed model simulations and formal analysis; AO and SdJ led the writing of the manuscript. All authors contributed critically to the drafts and gave final approval for publication.

\section*{Acknowledgements}
This work was supported by the Millennium Scientific Initiative ICM from Ministry of Economy, Development and Tourism of Chile to the Center for Marine Conservation {[}CCM RC 1300004{]}; AO was supported by the National Fund for Scientific and Technological Development, FONDECYT with a post-doctoral grant {[}grant number: 3150425{]} and by H2020 Marie Skłodowska-Curie Actions {[}746361{]}. KD Was supported by the Australian Research Council Centre of Excellence for Environmental Decisions {[}CE11001000104{]}, funded by the Australian Government. Additional funding was provided by Fondecyt \# 1160289 to SAN. Funding for the development of HYCOM has been provided by the National Ocean Partnership Program and the Office of Naval Research. Data assimilative products using HYCOM are funded by the U.S. Navy. Computer time was made available by the DoD High Performance Computing Modernization Program. The output is publicly available at \url{http://hycom.org/}

\bibliography{extracted}

% \end{doublespacing}	
\end{document}